# The radiation stability of glycine in solid CO₂ - *in situ* laboratory measurements with applications to Mars

Perry A. Gerakines and Reggie L. Hudson

*Astrochemistry Laboratory, Code 691*
*NASA Goddard Space Flight Center*
*Greenbelt, MD 20771*

perry.a.gerakines@nasa.gov
http://science.gsfc.nasa.gov/691/cosmicice

## Abstract

The detection of biologically important, organic molecules on Mars is an important goal that may soon be reached. However, the current small number of organic detections at the Martian surface may be due to the harsh UV and radiation conditions there. It seems likely that a successful search will require probing the subsurface of Mars, where penetrating cosmic rays and Solar energetic particles dominate the radiation environment, with an influence that weakens with depth. Toward the goal of understanding the survival of organic molecules in cold radiation-rich environments on Mars, we present new kinetics data on the radiolytic destruction of glycine diluted in frozen carbon dioxide. Rate constants were measured *in situ* with infrared spectroscopy, without additional sample manipulation, for irradiations at 25, 50, and 75 K with 0.8-MeV protons. The resulting half-lives for glycine in $CO_2$-ice are compared to previous results for glycine in $H_2O$-ice and show that glycine in $CO_2$-ice is much less stable in a radiation environment, with destruction rate constants $\sim$ 20-40 times higher than glycine in $H_2O$-ice. Extrapolation of these results to conditions in the Martian subsurface results in half-lives estimated to be less than 100-200 million years even at depths of a few meters.

## 1. Introduction

Many planetary bodies of astrobiological interest, such as Mars, are exposed to harsh incident radiation, which will influence the times that molecules can survive on them. Some or all of these bodies may well contain biologically-important organic molecules, some may even have supported life at some point in their history, and some may support life today. However, in the case of Mars, the Viking instruments were designed to search for evidence of life on the Martian surface (e.g., Anderson et al., 1972), but no organics were definitively detected (although a reanalysis suggests organic molecules may have been overlooked; see Navarro-González et al., 2010). Recently, chlorinated benzene was identified by the SAM instrument on the Curiosity Rover in samples drilled from rocks in Yellowknife Bay on Mars (Glavin et al., 2015), a sign that organic molecules may indeed survive just beneath the surface there. During flybys of Europa, near-infrared spectra from the Galileo-NIMS instrument showed no clear indications of organic molecules in surface ices (Carlson et al., 2009), although these remote observations had significant inherent limitations due to the harsh radiation environment around

Jupiter. Despite the recent upturn of results for one of these bodies, the history of non-detections for the surfaces of Mars and Europa are surprising because even if there is no endogenous organic material present, one expects a significant number of exogenous molecules delivered by meteorites, comets, and interplanetary dust particles. – For example, it has long been known that amino acids are present in the organic components of carbonaceous meteorites such as Murchison (Cronin et al., 1979), and more recently, organics have been detected in the tracks of cometary dust particles collected by the Stardust spacecraft as it flew through the coma of comet 81P/Wild 2. See Sandford et al. (2006). – One reason for the lack of organics found on both Europa and Mars is radiation-driven molecular destruction. Successful searches for organic molecules on these worlds likely will require sampling their subsurfaces, where organics may be frozen in ices dominated by either $H_2O$ or $CO_2$, which provide some protection from ionizing radiation. On Mars, solid $CO_2$ is observed in polar regions during Martian winter, and may persist throughout the summer in the southern polar region beneath the surface (Malin et al., 2001; Phillips et al., 2011), where organics and $H_2O$ also are likely to be embedded. Subsurface $H_2O$-ice on Mars may exist at various depths and experience seasonal temperature variations from 150 to 200 K (see, for example, thermal modeling by Bandfield, 2007). Turning to Europa, its surface consists of an $H_2O$-dominated ice shell at a temperature of $\sim$ 100-130 K (Carlson et al., 2009), whose chemistry is driven primarily by the bombardment by magnetospheric charged particles (keV-MeV electrons, especially in the trailing hemisphere; Paranicas et al., 2009).

The radiation doses received by molecules on planetary bodies depend on many factors and are extremely sensitive to depth beneath an exposed surface. For Mars, UV photons dominate at the near-surface (depths less than a few μm), whereas galactic cosmic rays and Solar energetic particles (mainly protons) dominate the subsurface particle-radiation environment. The MSL-RAD instrument recently measured the surface dose rate to be about 0.08 Gy yr$^{-1}$ (Hassler et al., 2014)**,** which is estimated to decrease with depth according to the content of the overlying material by factors of 5-10 to a depth of a few meters (also see Dartnell et al., 2007).

To accurately simulate and predict radiation-chemical behavior of organic molecules on Mars and other planetary bodies, kinetics data at large dilutions must be measured or extrapolated from existing data since accurate *a priori*, *ab initio* calculations are not feasible. Factors such as reactions of the surrounding material, for example an ice, or its radiation-chemical products are important and can be studied through reaction kinetics. Along these lines, in previous studies by our group we measured the radiation-chemical kinetics of glycine, the simplest amino acid, in $H_2O$-dominated ice mixtures relevant to icy planetary environments, with a focus on glycine's survival in the Martian subsurface (Gerakines et al., 2012; Gerakines and Hudson, 2013). An important observation was that rate constants and half-lives for the radiolytic destruction of glycine depend both on temperature and glycine concentration in $H_2O$-ice.

Outside of our own results, reports of the radiolytic decay of glycine are either based on analyses performed at room temperature or with glycine in an undiluted crystalline form. Moreover, most experiments in the literature were not performed at temperatures or with compositions appropriate to the surface of Mars or other Solar-System bodies, and the irradiated glycine samples were measured after significant sample handling and possible chemical alteration. For example, a commonly cited result is that of Kminek and Bada (2006), who gamma-irradiated dry glycine powder in sealed vials. After irradiation, the vials were opened and the contents dissolved in water for room-temperature liquid-chromatographic analysis to determine the loss of glycine, clearly compromising the details of the irradiation chemistry. Pilling et al. (2013) used infrared spectroscopy to measure the destruction rates of glycine polymorphs after exposure to 1-MeV protons, with most measurements again being made at room temperature. The $\alpha$ crystalline form of glycine was found to be the more susceptible to radiation-induced destruction, with the $\beta$ form being destroyed five times slower. These results agree with our own work (Gerakines et al., 2012).

Low-temperature (~10 - 30 K) glycine destruction also has been studied by other groups, some using *in-situ* methods. Pernet et al. (2013) used near-edge X-ray absorption spectroscopy to study the X-ray induced destruction rates of glycine in the absence and presence of $H_2O$-ice at 30 K. They concluded that dilution in $H_2O$-ice did not affect glycine's destruction rate, and predicted glycine half-lives of about 1 year at 1 AU from the Sun. However, their measurements of radiation dose were restricted to the exposure time for each sample and no measurements of *absorbed* radiation doses could be made, making half-life comparisons for wet and dry samples difficult. Johnson et al. (2012) measured glycine destruction at 18 K in an argon matrix (Ar:gly ratio ~10,000:1) exposed to UV photons from three sources with different emission lines. Results showed that glycine was easily destroyed by UV photons, such as would be present at the surface of Mars or Europa, with half-lives of a few years. Orzechowska et al. (2007) also used UV photolysis to measure the photo-destruction of glycine, but in mm-thick $H_2O$-ice samples at 100 K. Their analyses used room-temperature liquid chromatography, again requiring the melting of the ice samples after photolysis. Even earlier work is available on the photo-destruction of amino acids at 12 K, for applications to interstellar chemistry, but there too only incident fluences (photons $cm^{-2}$) were measured, not absorbed energies (Ehrenfreund et al., 2001).

In short, we know of no published *in-situ* determinations of low-temperature radiolytic destruction of glycine or any other amino acids mixed with $H_2O$-ice other than our earlier work (Gerakines et al., 2012; Gerakines and Hudson, 2013). For the case of amino acids, such as glycine, in $CO_2$-ice the literature is even sparser. Although $CO_2$ is a major constituent of ices on cold planetary bodies, no radiation-chemical results for mixtures of $CO_2$ + glycine are in the literature. Therefore, in the present study we report laboratory measurements of the survival of glycine diluted in $CO_2$-ice at several concentrations ($CO_2$:glycine = 75:1, 150:1, 190:1, 250:1, and 380:1) and at three temperatures (25, 50, and 75 K). For each sample, we measured the destruction yield ($G$-value), destruction rate constant, and half-life dose of glycine due to irradiation by 0.8-MeV protons. All measurements were made *in situ* at the temperature of irradiation using infrared (IR) spectroscopy. While the temperatures and concentrations were limited by experimental constraints and do not mimic those on Mars or Europa, observed trends in the laboratory results can be extrapolated to relevant planetary environments.

## 2. Experimental details

### 2.1. Sample preparation

The experimental system in the Cosmic Ice Laboratory at the NASA Goddard Space Flight Center has been described in detail (Hudson and Moore, 1999), and the details of the most current set-up can be found in our previous study on amino acids (Gerakines et al., 2012). In summary, our system consists of a high-vacuum chamber (P = $5 \times 10^{-7}$ torr at 300 K) mated to the beam line of a Van de Graaff accelerator and to an IR spectrometer (Nicolet Nexus 670 FT-IR). A polished aluminum substrate (area $\approx$ 5 cm$^2$) is mounted inside the chamber on the end of the cold finger of a closed-cycle helium cryostat (Air Products Displex DE-204) capable of cooling to a minimum of $\sim$ 15 K. The substrate's temperature is monitored by a silicon diode sensor and can be adjusted up to 300 K using a heater located at the top of the substrate holder. This same substrate is positioned so that the IR spectrometer's beam is reflected from the substrate's surface at a near-normal angle ($\sim$ 5°) and directed onto an HgCdTe (MCT) detector. With an ice sample on the metal substrate, the IR beam passes through the sample before and after reflection off the underlying metal surface. The substrate is fully rotatable through 360° to face the IR spectrometer, the Van de Graaff accelerator, and other components.

Samples were prepared as follows. A custom-built Knudsen-type sublimation oven, attached to one port of the vacuum chamber, was used to sublime glycine into the chamber and produce micrometer-thick films on the cold substrate (for more details about the oven see Gerakines et al., 2012). With about 50 mg of glycine in the sublimation oven, it was heated (to 160°C) until a sufficient amount of glycine vapor was produced to create a sample. At that point, $CO_2$ vapor was released into the vacuum chamber in front of the oven using a metered leak valve, which was calibrated as described by Gerakines et al. (2012) to achieve the desired $CO_2$-to-glycine ratio for the sample to be studied. With both gases flowing, the cold (25 K) substrate then was rotated to face the oven, and the sample's growth was measured by monitoring the interference fringes of a 650-nm laser beam reflected from the sample and substrate surfaces at an incidence angle of 10°. The sample growth rates in this study varied according to the final desired glycine dilution and were in the range of 8 - 40 μm hr$^{-1}$, with final film thicknesses of about 3 μm and resulting $CO_2$-to-glycine ratios from 75:1 to 380:1. All $CO_2$ + glycine samples were grown at 25 K and heated at 1 K min$^{-1}$ to the irradiation temperatures stated below. All IR spectra were measured using 150 scans from 5000 to 650 cm$^{-1}$ with a resolution of 2 cm$^{-1}$. The compounds, sources, and purities used in our experiments were as follows: glycine, Sigma Chemical Co., 99% purity; $CO_2$, Aldrich Chemical Co., 99.99% purity.

*2.2. Radiation doses*

All samples were irradiated with a beam of 0.8 MeV protons at a current of $\sim 10^{-7}$ A. For each radiation dose, the ice first was rotated to face the proton beam from the accelerator and then turned 180° so as to face the IR beam of the spectrometer to record the sample's spectrum. The dose absorbed is given by

$$\text{Dose [eV g}^{-1}\text{]} = SF \qquad (1)$$

where $S$ is the proton stopping power (in eV cm$^2$ g$^{-1}$ p+$^{-1}$) and $F$ is the incident proton fluence (in p+ cm$^{-2}$). Doses in the SI units of gray (Gy) equal $SF \times (1.60 \times 10^{-16}$ Gy g eV$^{-1}$), where 1 Gy = 1 joule absorbed / kg of sample. For conversions to other units see, for example, Gerakines and Hudson (2013).

The stopping power for 0.8-MeV protons in each sample was calculated using the SRIM software package (Ziegler et al., 2010). For glycine, we assumed a visible refractive index $n_{\text{vis}} =$ 1.46 and a mass density $\rho = 1.61$ g cm$^{-3}$ (Weast et al., 1984). For $CO_2$, we used $n_{vis} = 1.25$ and $\rho = 1.12$ g cm$^{-3}$ as measured at 25 K by Satorre et al. (2008). We then assumed a weighted average for $n_{vis}$ and $\rho$ for mixtures based on the $CO_2$:gly ratio. Since all mixtures were dominated by $CO_2$ by at least 75 to 1, the stopping power, visible refractive index, and assumed density of each sample were always within 1% of $S = 2.41 \times 10^8$ eV cm$^2$ g$^{-1}$ p+$^{-1}$, $n_{vis} = 1.25$, and $\rho = 1.12$ g cm$^{-3}$. This means that for these $CO_2$-dominated ices an exposure to $1 \times 10^{13}$ p+ cm$^{-2}$ corresponds to an absorbed dose of 0.39 MGy.

## 3. Results

All IR spectra of $CO_2$ + gly mixtures contained sharp glycine features at 3510, 1775, 1140, and 825 cm$^{-1}$. Although glycine in both crystalline form and in $H_2O$-ice can appear as a zwitterion (Gerakines et al., 2012; Portugal et al., 2014), the low deposition temperatures and high dilutions ($\geq 75$:1) we used favor the formation of glycine in its non-zwitterionic form. Figure 1 contains the IR spectrum of a $CO_2$ + glycine (75:1) ice sample after deposition at 25 K (bottom trace), where glycine's IR absorptions are marked with asterisks (the composition and temperature of the ice used for this figure were those most clearly displaying the glycine features and those of the irradiation products). We note that the IR spectrum of glycine mixed with $CO_2$ ice at 25 K was presented recently by Maté et al. (2011), but that their spectrum differs from what was observed by us (Fig. 1) in that all of the glycine features reported by Maté et al. were much broader. However, they admit that the glycine concentration in their samples was highly uncertain due to a lack of intrinsic (absolute) IR band strengths with which to calculate ice compositions from spectra.

Although the spectroscopy and chemistry of irradiated glycine are intrinsically interesting, the present study focuses instead on the kinetics of glycine destruction. Figure 1 shows that upon proton irradiation all glycine IR features decreased in intensity for $CO_2$ + gly (75:1) at three irradiation doses at 25 K. Changes to the $CO_2$ absorptions were expected to be small at these dose levels and were indeed not measureable. Glycine decomposition products, such as methylamine, also were not observed in the IR. However, irradiation products of the dilutant $CO_2$ were detected, such as the CO absorption feature at 2140 cm$^{-1}$ and $CO_3$ absorptions at 2045, 1880, and 973 cm$^{-1}$ (Moll et al., 1966). Of note is that no evidence was found for the formation of CH$_4$, which was recently detected as a trace gas in the martian atmosphere by MSL-SAM (Webster et al., 2014). The 2160 cm$^{-1}$ feature of the OCN$^-$ anion also was not observed in the irradiated ices.

### 3.1. Destruction kinetics

To determine the loss of glycine due to an irradiation, we monitored the area underneath the glycine IR absorption at 1140 cm$^{-1}$, which is relatively uncontaminated by overlapping features and can be taken as proportional to the number of glycine molecules ($N_{gly}$) in a sample. For all ices studied, the area of this IR band decreased exponentially with the total dose (proton fluence, $F$) and followed first-order kinetics. Figure 2 shows typical decay curves for three $CO_2$ + glycine ices at 75 K, with the IR band areas ($\mathcal{A}$) at 1140 cm$^{-1}$ normalized to the initial area ($\mathcal{A}_0$).

For our kinetics analysis, we assumed the form

$$\text{glycine} \rightleftharpoons \text{products} \qquad (2)$$

for amino-acid destruction, with both the forward and reverse reactions taken as first order (Gerakines and Hudson, 2013). This led to an appropriate fit of the glycine decay data to the equation

$$(\mathcal{A}/\mathcal{A}_0) = a\,e^{-bF} + c \qquad (3)$$

where $a$, $b$, and $c$ are independent parameters. Also note that since IR band areas in these samples are proportional to the number of absorbing molecules, equation (3) represents the fraction of glycine molecules remaining in the ice after fluence $F$. Parameter $a$ represents the fractional loss of glycine after long irradiation times, $b$ is the sum of the forward and backward rate constants in equation (2), in units of $cm^2\ p+^{-1}$, and $c$ is the equilibrium fraction of glycine. In Fig. 2, curve fits in the form of equation (3) are shown for $CO_2$ + gly mixtures irradiated at 75 K. Figure 2 also shows the glycine destruction at 75 K as a function of dose in units of MGy and total energy absorbed (in eV). The best-fitting parameters for all of the samples are listed in Table 1. Note that although we did not place any constraints on the parameters, the sum $a + c$ is approximately equal to 1 (as expected) within the listed uncertainties for each sample.

The destruction rate constant $k$ can be derived from the curve-fit parameters in Table 1. At low doses, when $bF \ll 1$, the right-hand side of equation (3) can be approximated by $1 - abF$ (since $a + c \approx 1$). This is the same form as a first-order destruction rate law approximated at early time steps by zeroth-order kinetics, where the time variable is replaced with $F$ and the rate constant with $ab$ (in units of $cm^2\ p+^{-1}$). Expressed in units of dose $D$ (given in eV $g^{-1}$ or in MGy), the fractional loss of glycine is initially given by $kD$ (where $k = ab/S$), and so the destruction rate constant $k$ is a measure, per unit dose, of the glycine lost relative to the initial number of glycine molecules.

## 4. Discussion

### 4.1. Destruction rate constants and yields

For each of the twelve ices of Table 1, we used the curve-fit parameters $a$, $b$, and $c$ to determine both the corresponding forward rate constant $k$, as already described, and the initial glycine destruction yield, $G$(-gly), defined as the number of glycine molecules destroyed per 100 eV absorbed by the sample at low doses (see Gerakines et al., 2012). The resulting $k$ and $G$(-gly) values are listed in Table 2 along with corresponding half-life doses. We stress that $G$, strictly speaking, describes a radiation-chemical yield only in the early stages of a reaction. Although $G$ has long been used by radiation chemists (Burton, 1952), it has limited value for extrapolations over astronomical time periods for which rate constants such as $k$ can be more applicable.

Rate constants for glycine destruction are plotted in Fig. 3 against the initial $CO_2$:gly ratio of our samples. A slight rise can be seen with increasing glycine dilution, implying that greater numbers of $CO_2$ molecules produce greater fractional destruction rates of glycine per unit dose.

Figure 4 plots $k$ against the temperature at which each ice sample was irradiated. Due to the low sublimation temperature of $CO_2$ in vacuum, our data are restricted to $T \leq 75$ K but, although somewhat limited, do show an increase in the destruction rate constant with irradiation temperature.

The trends observed in Figs. 3 and 4 can be explained by indirect, chemical action of the irradiation through reactive species produced from the $CO_2$ ice itself, which become more abundant relative to glycine at higher dilutions and more mobile at higher temperatures. As has long been known (e.g., Moll et al., 1966), irradiation of solid $CO_2$ produces O atoms which react with each other to form ozone and with $CO_2$ to form higher oxides, such as $CO_3$. The glycine component of our $CO_2$ + glycine mixtures also will be affected by these O atoms. The increase in glycine destruction with irradiation temperature seen in Fig. 4 could be linked to the mobility of any of these reactive atoms or molecules within the $CO_2$ ice.

### 4.2. Comparisons with glycine in $H_2O$-ice

Since $k$ depends strongly on the overall composition of the glycine-containing sample, it is appropriate to compare the present results with those from our earlier study (Gerakines and Hudson, 2013). Figures 3 and 4 show our $CO_2$ results and also the destruction rate constants of glycine in $H_2O$-ice from Gerakines and Hudson (2013) plotted versus $CO_2$:gly ratio (Fig. 3) and temperature (Fig. 4). The difference in the rate constants for the two types of ice is striking. Figure 5 summarizes our results in a different way, showing decay curves for $H_2O$ + gly (300:1) irradiated at 100 K and $CO_2$ + gly (380:1) irradiated at 75 K. The differences in the surviving fractions of glycine at each dose are obvious. The magnitude of the initial slope of each decay curve corresponds to the value of $k$ for destruction. As illustrated in Fig. 5 and indicated by the $k$ values, glycine is destroyed about 40 times faster in $CO_2$-ice than in $H_2O$-ice.

A minor contribution to the relatively higher destruction rates for glycine in $CO_2$ is the shielding provided by the solid from the incident radiation, and different molecules provide different levels of shielding. The mass stopping power (the amount of energy absorbed per unit surface density) for $CO_2$ is about 20% lower than that for $H_2O$ ($2.4 \times 10^8$ versus $2.9 \times 10^8$ eV cm$^2$ g$^{-1}$ p+$^{-1}$). This means that less energy is absorbed by $CO_2$-ice than by $H_2O$-ice, and therefore the former provides less direct protection to the glycine component. The formation of reactive radiolysis products such as O atoms and $O_3$ molecules may account for the much higher glycine destruction rates in $CO_2$-ice over those observed in $H_2O$-ice.

Another important aspect to note is the fraction of glycine that survives after very long exposures to radiation (the equilibrium fraction), given by the curve-fit parameter $c$ in equation (3) and Table 1. In each of the $CO_2$ + gly samples studied here, $c$ is 1-4%, with an uncertainty of about 1%. In our $H_2O$ + gly mixtures, we found significantly higher values, up to about 50 - 60%. This shows that not only does glycine have a much greater destruction rate in $CO_2$ ice, but also that very little, if any, of the initial glycine remains after long exposure times.

### 4.3. Implications for glycine on Mars or in comets

In Table 3 we list dose rates expected for various depths on Mars and the expected half-lives of glycine in $H_2O$-ice and in $CO_2$-ice. For these calculations we used data from the highest temperature at which we could study $CO_2$-ice and a similar value for $H_2O$-ice, and for the most dilute glycine samples we have studied.  Note that our half-life estimates are based on the dose rates of Hassler et al. (2014), which are from measurements at the surface of Mars by the Curiosity rover's RAD instrument and which are smaller than the calculated results from atmospheric models (Dartnell et al., 2007) by about a factor of 2.  Table 3 and Figs. 3 and 4 show that the half-life doses for glycine in $CO_2$-ice are much lower than for glycine mixed with frozen $H_2O$, which leads to much shorter survival times on or beneath the surface of Mars when glycine is mixed with $CO_2$.  Overall, the trends of Table 2 and Fig. 4 suggest a rising rate constant for the radiolytic destruction of glycine in frozen $CO_2$ as temperature increases, but little change in $k$ for glycine in $H_2O$ ice.  Therefore, a cautious extrapolation to the warmer temperatures of Mars suggests that the differences between glycine in $CO_2$-ice and $H_2O$-ice shown in Figs. 4 and 5 may be much greater on Mars than displayed in our experiments.

Conclusions similar to those drawn for glycine on Mars also apply to Europa and other icy satellites and to comets.  For the specific case of comets, Ryan and Draganic (1986) estimated the total dose received over $4.6 \times 10^9$ years by an Oort-Cloud body from cosmic-ray irradiation at various depths beneath the cometary surface.  Based on their Fig. 3, doses can be as high as 330 MGy at a depth of 10 cm, falling to about 133 MGy at a 1-m depth, 17 MGy at 5 m, and 3 MGy at 10 m.  According to our Fig. 5, no significant amount of glycine could exist in a comet's $CO_2$ ices closer to the surface than about 10 meters.  However, a large fraction of glycine could survive trapped in $H_2O$-ice at depths above 5 meters.

## 5. Summary and conclusions

Destruction rate constants were measured for glycine in ices ranging in composition from $CO_2$:gly = 75:1 to 380:1 after irradiation with 0.8 MeV protons, with all measurements being made *in situ* with IR spectroscopy at 25 - 75 K.  The half-life of glycine was found to depend on the ice's composition and the irradiation temperature.  Observed decays indicated very low equilibrium (residual) fractions of glycine (1-4 % of the initial glycine molecules) as a result of irradiation, with destruction rate constants that are 20-40 times larger than those for glycine in an $H_2O$ ice.

Laboratory conditions were limited to temperatures below 75 K and to high-vacuum pressures, while $CO_2$ ice on Mars exists at a temperature of $\sim$ 150 K, with pressure of a few torr.  Given the observed trend in Figure 4, the higher martian temperature implies a 25-50 % faster glycine destruction rate, while the increase in gas pressure should not have a significant effect on the glycine destruction measured here.

On the surface of Mars, glycine mixed with $CO_2$-ice should have a half-life of a few million years, and if glycine persists on Mars in $CO_2$-ice for more than a few hundred million years then it must be in areas deeper than a few meters, where it would be shielded from cosmic rays.  This study cannot predict the presence of glycine on Mars, only its persistence.  However, from our data it is clear that given a choice of searching for glycine on Mars in either $H_2O$ or $CO_2$ ices, glycine is much more likely to be found in $H_2O$-ice.

**Acknowledgments**

We acknowledge the continuing support of the NASA Astrobiology Institute (NAI) and the Goddard Center for Astrobiology (GCA).  In addition, we thank Steve Brown, Tom Ward, and Eugene Gerashchenko, members of the Radiation Effects Facility at NASA Goddard Space Flight Center, for operation of the proton accelerator.

Table 1.  Parameters for glycine radiolytic decay in solid $CO_2$.

| Composition $CO_2$ : glycine | T [K] | Curve-fit parameters* | | |
| | | $a$ | $b$ [$10^{-14}$ cm$^2$ p+$^{-1}$] | $c$ |
|---|---|---|---|---|
| 75:1 | 25 | 0.99 ± 0.01 | 3.47 ± 0.05 | 0.01 ± 0.01 |
| 75:1 | 50 | 0.97 ± 0.01 | 3.68 ± 0.10 | 0.02 ± 0.01 |
| 75:1 | 60 | 0.97 ± 0.01 | 4.00 ± 0.10 | 0.02 ± 0.01 |
| 150:1 | 25 | 0.96 ± 0.01 | 4.70 ± 0.17 | 0.03 ± 0.01 |
| 150:1 | 50 | 0.95 ± 0.01 | 4.83 ± 0.10 | 0.04 ± 0.01 |
| 150:1 | 75 | 0.95 ± 0.01 | 5.53 ± 0.18 | 0.04 ± 0.01 |
| 250:1 | 25 | 0.98 ± 0.01 | 5.47 ± 0.10 | 0.02 ± 0.01 |
| 250:1 | 50 | 0.97 ± 0.01 | 5.48 ± 0.15 | 0.02 ± 0.01 |
| 250:1 | 75 | 0.97 ± 0.01 | 6.37 ± 0.20 | 0.02 ± 0.01 |
| 380:1 | 25 | 0.98 ± 0.01 | 5.32 ± 0.01 | 0.02 ± 0.01 |
| 380:1 | 50 | 0.99 ± 0.01 | 6.83 ± 0.13 | 0.01 ± 0.01 |
| 380:1 | 75 | 0.98 ± 0.02 | 7.40 ± 0.23 | 0.02 ± 0.01 |

* All fits are in the form of $(\mathcal{A}/\mathcal{A}_0) = a\, e^{-bF} + c$ (equation 3).

Table 2. Derived glycine destruction properties for twelve ices.

| Composition CO$_2$:glycine | T [K] | $G$(−gly) [a] [molec (100 eV)$^{-1}$] | Destruction rate constant, $k$ [b] | | Half-life dose [c] | |
|---|---|---|---|---|---|---|
| | | | [10$^{-14}$ cm$^2$ p+$^{-1}$] | [MGy$^{-1}$] | [10$^{13}$ p+ cm$^{-2}$] | [MGy] |
| 75:1 | 25 | 2.60 ± 0.04 | 3.44 ± 0.06 | 0.89 ± 0.01 | 2.00 ± 0.04 | 0.77 ± 0.02 |
| 75:1 | 50 | 2.70 ± 0.08 | 3.57 ± 0.11 | 0.92 ± 0.03 | 1.91 ± 0.08 | 0.74 ± 0.03 |
| 75:1 | 60 | 2.87 ± 0.07 | 3.88 ± 0.09 | 1.01 ± 0.02 | 1.76 ± 0.05 | 0.68 ± 0.02 |
| 150:1 | 25 | 1.67 ± 0.06 | 4.52 ± 0.18 | 1.17 ± 0.05 | 1.54 ± 0.08 | 0.59 ± 0.03 |
| 150:1 | 50 | 1.75 ± 0.04 | 4.61 ± 0.11 | 1.19 ± 0.03 | 1.50 ± 0.04 | 0.58 ± 0.02 |
| 150:1 | 75 | 2.00 ± 0.07 | 5.27 ± 0.19 | 1.36 ± 0.05 | 1.30 ± 0.06 | 0.50 ± 0.02 |
| 250:1 | 25 | 1.21 ± 0.02 | 5.37 ± 0.10 | 1.39 ± 0.03 | 1.29 ± 0.03 | 0.50 ± 0.01 |
| 250:1 | 50 | 1.22 ± 0.04 | 5.31 ± 0.16 | 1.38 ± 0.04 | 1.30 ± 0.05 | 0.50 ± 0.02 |
| 250:1 | 75 | 1.42 ± 0.05 | 6.18 ± 0.21 | 1.60 ± 0.05 | 1.11 ± 0.05 | 0.43 ± 0.02 |
| 380:1 | 25 | 0.77 ± 0.01 | 5.20 ± 0.01 | 1.35 ± 0.02 | 1.34 ± 0.03 | 0.52 ± 0.01 |
| 380:1 | 50 | 1.00 ± 0.02 | 6.74 ± 0.14 | 1.75 ± 0.04 | 1.03 ± 0.03 | 0.40 ± 0.01 |
| 380:1 | 75 | 1.08 ± 0.04 | 7.28 ± 0.24 | 1.89 ± 0.06 | 0.96 ± 0.04 | 0.37 ± 0.02 |

[a] Number of glycine molecules destroyed per 100 eV absorbed by the sample.
[b] Forward rate constant in the chemical reaction given by equation (2).
[c] Energy dose required to reduce the initial glycine abundance by 50%.

Table 3. Dose rates and glycine half-lives under the surface of Mars.

| Depth / cm | Dose rate[a] / mGy yr$^{-1}$ | Glycine half-life / $10^6$ yr | |
| --- | --- | --- | --- |
| | | in $H_2O$-ice[b] | in $CO_2$-ice[c] |
| 0 | 76 | 190 | 5.7 |
| 10 | 96 | 150 | 3.8 |
| 100 | 36.4 | 401 | 10 |
| 200 | 8.7 | 1700 | 42 |
| 300 | 1.8 | 8100 | 205 |

[a] Hassler et al. (2014).

[b] $H_2O$ + gly (300:1) at 100 K with a half-life dose of 14.6 MGy (Gerakines and Hudson, 2013).

[c] The listed half-lives correspond to the most dilute sample in our data set, $CO_2$ + gly (380:1) irradiated at 75 K, with a half-life dose of 0.37 MGy (Table 2).

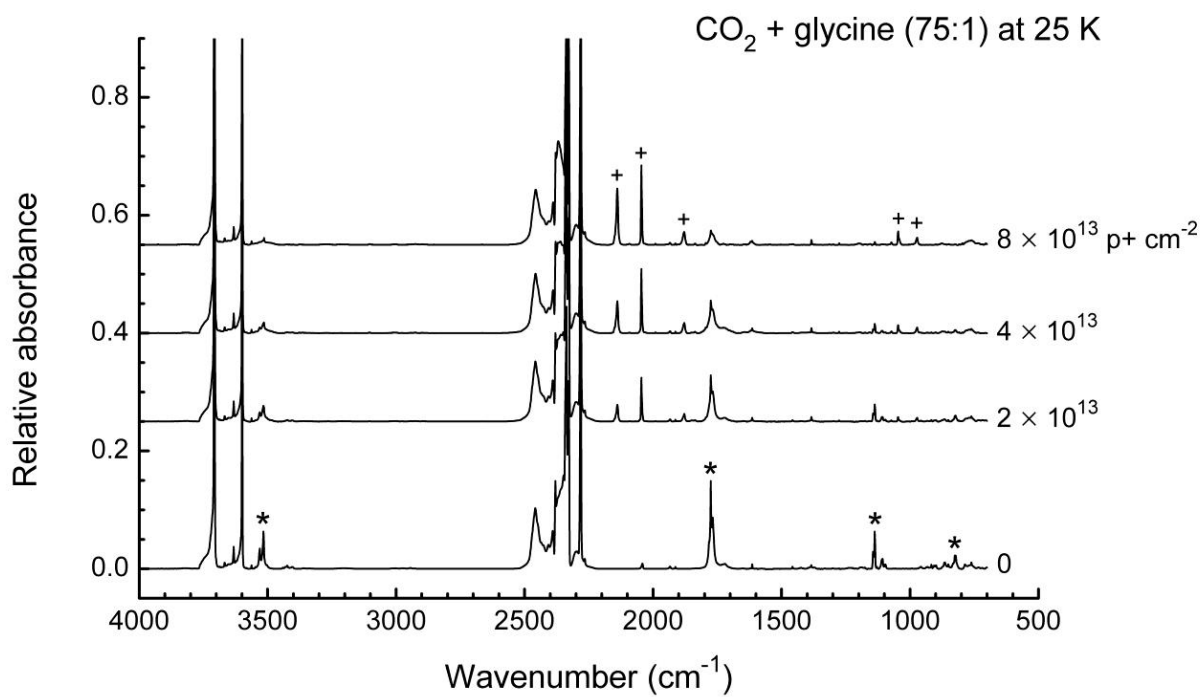

**Fig. 1.** The IR spectrum of a 3.1-μm thick sample composed initially of $CO_2$ + glycine (75:1) and irradiated at 25 K. The bottom trace is before irradiation, with spectra after successive doses from bottom to top. Glycine peaks are marked with asterisks in the bottom trace, and peaks from $CO_2$ irradiation products are indicated by "+" in the top trace. The total accumulated proton fluence after each irradiation is listed, with $1 \times 10^{13}$ p+ cm$^{-2}$ = 0.39 MGy.

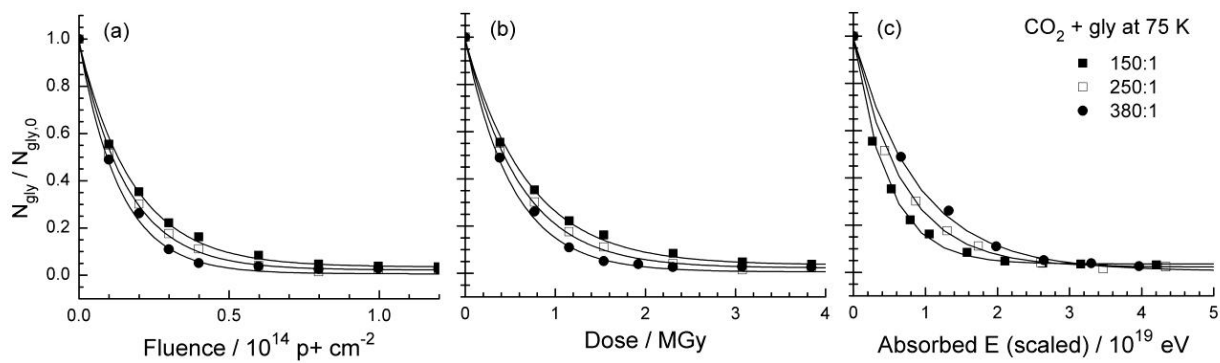

**Fig. 2.** Radiation-induced destruction of glycine in $CO_2$ + glycine ices at 75 K with initial $CO_2$:gly ratios of 150:1, 250:1, and 380:1. In (c), the absorbed energy for each sample has been scaled to match that for the same mixture with an initial number of glycine molecules equal to $10^{17}$. The solid lines connecting the data points in each panel are offset exponential fits in the form of equation (3) converted for the appropriate x-axes shown.

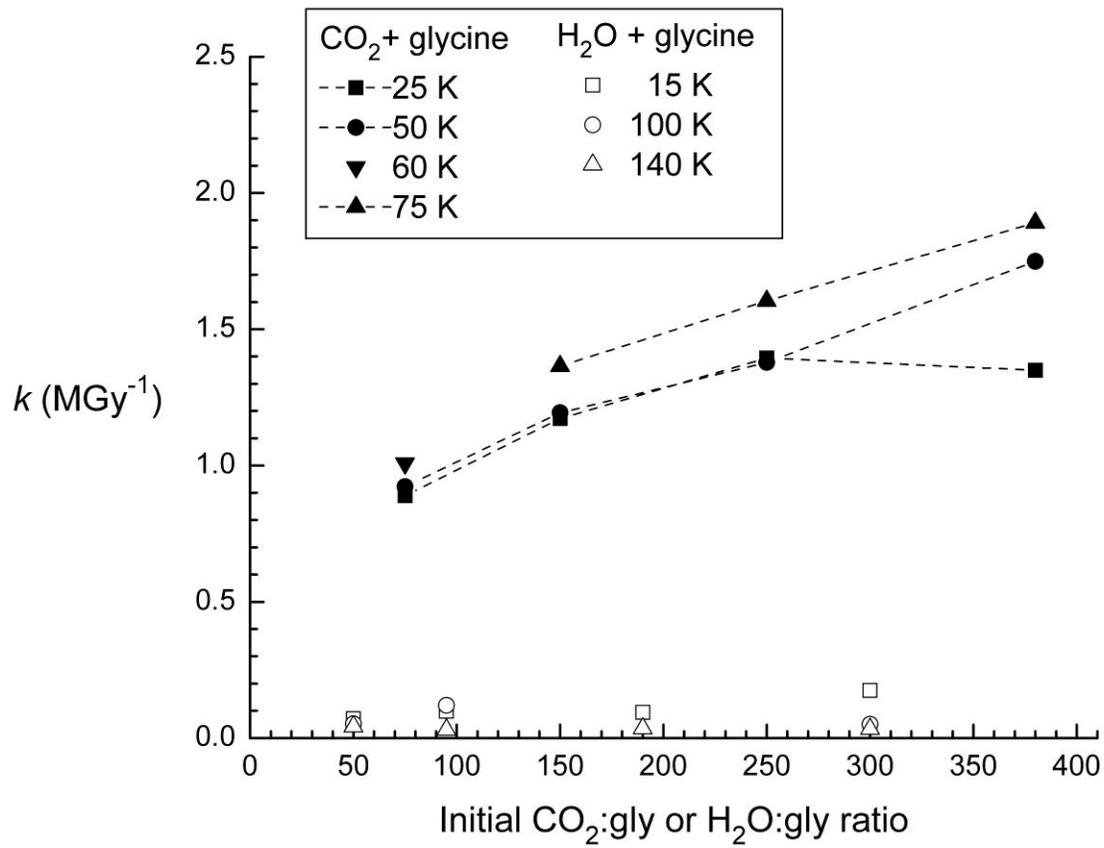

**Fig. 3.** Glycine destruction rate constants (in MGy$^{-1}$) for $CO_2$ + glycine samples with different compositions and corresponding data from $H_2O$ + glycine ices (Gerakines and Hudson, 2013).

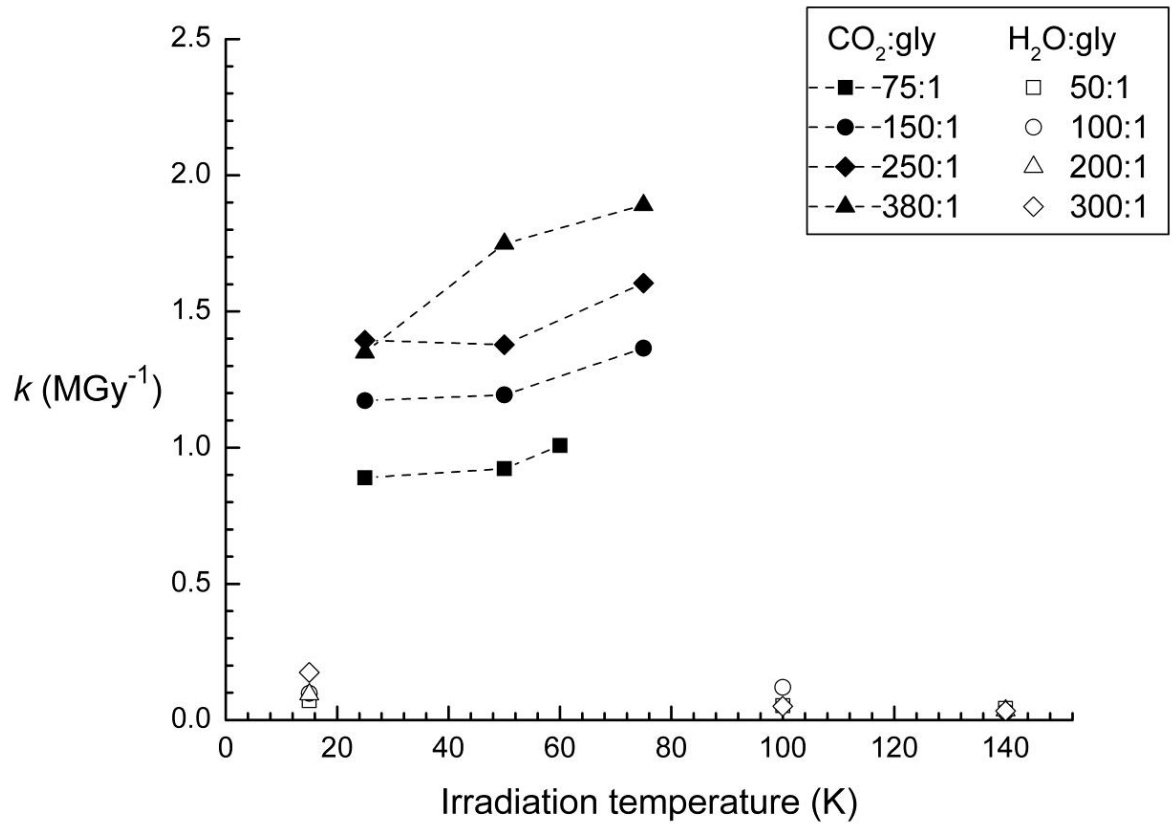

**Fig. 4.** Glycine destruction rate constants (in $MGy^{-1}$) for $CO_2$ + glycine samples irradiated at different temperatures compared to results from $H_2O$ + glycine ices (Gerakines and Hudson, 2013).

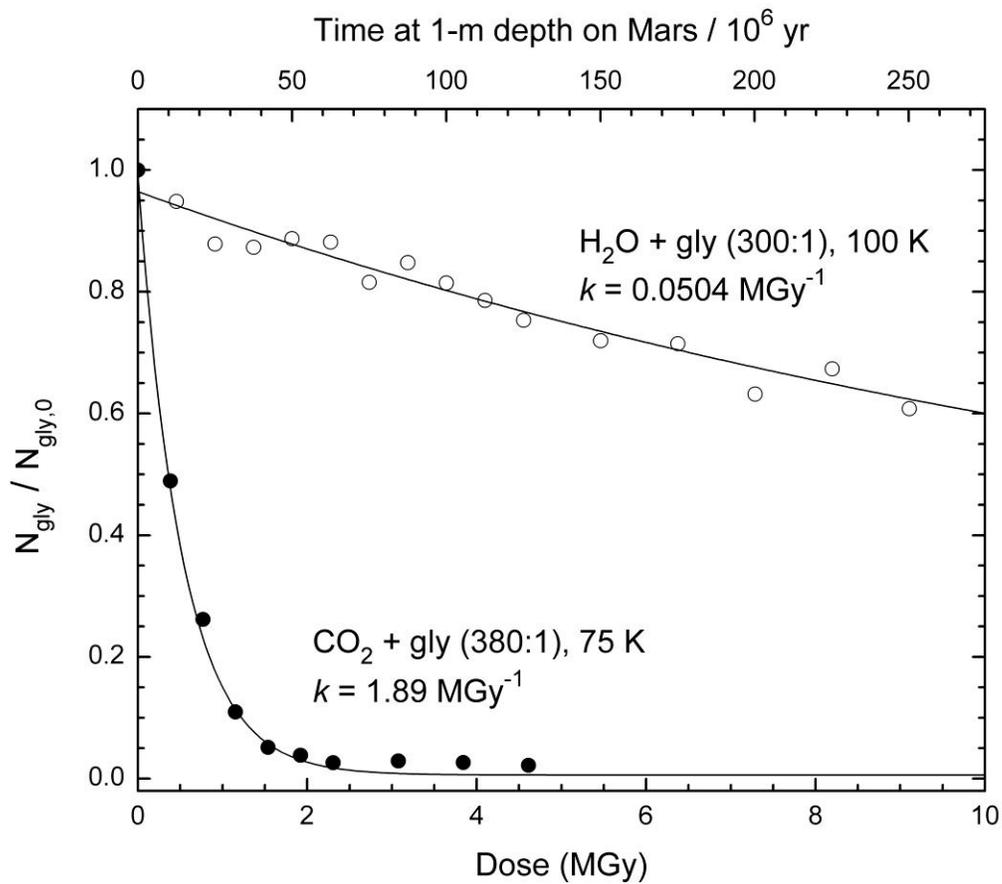

**Fig. 5.** A comparison of the surviving fraction of glycine molecules versus dose (in MGy) for glycine diluted and irradiated in two different ices, $H_2O$ and $CO_2$. The top axis gives corresponding times at a depth of 1 m on Mars based on the dose rate of 36.4 mGy yr$^{-1}$ (Hassler et al., 2014).